\documentclass[mathleft
]{an}
\usepackage{graphicx}
\usepackage{times}
\begin{document}

\Pagespan{262}{265}
\Yearpublication{2011}%
\Yearsubmission{2011}%
\Month{02}%
\Volume{332}%
\Issue{3}%
\DOI{10.1002/asna.201111531 }%

\title{Studying the SN-GRB Connection with X-shooter: \\
the GRB 100316D / SN 2010bh case\,\thanks{
Based on observations collected at the European Southern Obser- 
vatory, ESO, the VLT/Kueyen telescope, Paranal, Chile, proposal codes 084.D-0265 and 085.D-0701, P.I. S.Benetti and
 084.A-0260 and 085.A-0009, P.I. J.Fynbo.}}

\author{F. Bufano\inst{1}\fnmsep\thanks{Corresponding author:
  \email{filomena.bufano@oact.inaf.it}\newline}
\and S. Benetti \inst{2} \and J.  Sollerman \inst{3} \and E. Pian \inst{4,5} \and G. Cupani \inst{4}
}
\titlerunning{Studying SN-GRB Connection with X-Shooter}
\authorrunning{F. Bufano et al.}
\institute{
INAF - Osservatorio Astronomico di Catania, Via S.Sofia 78, I-95123, Catania, Italy 
\and 
INAF - Osservatorio Astronomico di Padova, Vicolo dell'Osservatorio 5, I-35122, Padova, Italy
\and
The Oskar Klein Centre, Department of Astronomy, AlbaNova,	SE-106 91, Stockholm, Sweden
\and
INAF - Osservatorio Astronomico di Trieste, Via G.B. Tiepolo 11, I-34143, Trieste, Italy
\and 
Scuola Normale Superiore, Piazza dei Cavalieri 7, I-56126, Pisa, Italy		
}

\received{2011 Feb 02}
\accepted{2011 Feb 07}
\publonline{2011 Mar 17}


\abstract{ 
During the last ten years, observations of long-duration gamma-ray bursts brought to the conclusion that at least a fraction of them is associated with bright supernovae of type Ib/c.
In this talk, after a short review on the previously observed GRB-SN connection cases, we present the recent case of GRB 100316/ SN 2010bh. 
In particular, during the observational campaign of SN 2010bh, a pivotal role was played by VLT/X-shooter, 
sampling with unique high quality data the spectral energy distribution of the early evolution phases from the UV to the K-band. }

\maketitle
\section{Introduction}
	During the last few years, the link between long-duration gamma-ray bursts 
	(GRBs) and type Ib/c core-collapse supernovae (SNe) 
	has been firmly established. In 1998,   we had with GRB 980425 and  SN 1998bw (Galama et al. 
	1998) the first clear case of the connection between GRBs and SNe. This connection was later confirmed by three further associations between 
	nearby (out to z = 0.17) GRBs and spectroscopically-confirmed SNe: GRB 030329/SN 2003dh (Hjorth et al. 2003; Stanek et al. 
	2003), GRB 031203/ SN 2003lw (Malesani et al. 2004) and GRB 060218/ SN 2006aj (Campana et al. 2006; Pian et al. 2006). 
	GRB-SN connection cases are rare, but represent a unique tool to understand the nature of their progenitor.
         In this paper, we present SN 2010bh (Wiersema et al. 2010; Bufano et al. 2010; Chorno\-ck et al. 2010a,b), the latest spectroscopically confirmed 
         SN associated with a nearby long-GRB, GRB 100316D (Stamatikos et al. 2010).\\
         During the observational campaign of SN 2010bh, a pivotal role was played by X-shooter, the new generation spectrograph mounted on the Unit Telescope 
         2 at the Very Large Telescope (VLT), that sampled with unique high quality da\-ta its spectral energy distribution (SED) from the UV to the K-band starting from early evolutionary phases.
   
\begin{table*}
\begin{center}
\caption{Main properties of the four nearby GRBs connected with spectroscopically confirmed SNe.}
\label{tab}
\begin{tabular}{cccccccccc}
\hline
&z&$T_{90}$&$E_{iso}$& $t_{max(V)}$& $E_k$ & $M_{Ni}$& $M_{ej}$ &$M_{MS} $ &Ref.\\
& & [s] & [erg] & [d]&[erg]& [$M_\odot$]& [$M_\odot$]& [$M_\odot$]&\\
\hline
GRB 980425/SN 1998bw&0.0085&23.3&10$^{48}$&17& $5 \times 10^{52}$& 0.38--0.48&11&40&1,2,3,5\\
GRB 030329/SN 2003dh&0.168&25&$1.7\times10^{52}$&10--13&$4\times 10^{52}$& 0.25--0.45& 8& 25--40&2,4,5 \\
GRB 031203/SN 2003lw&0.105&20&9--10$\times10^{49}$&20&$6\times10^{52}$&0.45--0.65&13&40--50&2,6,7\\
GRB 060218/SN 2006aj&0.033&2100&$6\times10^{49}$&11.4&$2\times10^{51}$&0.21&2&20--25&2,8,9\\
\hline
\end{tabular}
\end{center}
{\footnotesize(1) Galama et al. (1998);(2) Amati et al. (2007); (3) Iwamoto et al (1998); (4) Hjorth et al. 2003; (5) Mazzali et al. (2006b); (6) Malesani et al (2004); (7) Mazzali et al. (2006b); (8) Pian et al. (2006); (9) Mazzali et al (2006a)}
\end{table*}

\section {Previous GRB-SN connections in the nearby Universe}
	SN 1998bw was discovered as an optical transient in the error box of GRB 980425: the spectroscopic 
	and photometric evolution of this event resembled more that of a SN Ic rather than a typical GRB afterglow.  The chance of  a spatial coincidence of the two events was 
	estimated to be less than $\sim10^{-4}$ (Galama et al. 1998). In  particular, both SN 1998bw and GRB 980425 showed a peculiar behavior:  the GRB was 4 orders of magnitude 
	less energetic than a normal GRB ($E_{iso}\sim10^{48}$erg, see Table \ref{tab}), while the SN was the most luminous SN ever observed, characterized by a  peak luminosity of  
	$10^{43}$ erg s$^{-1}$ (Iwamoto et al. 1998),  10 times higher than a normal type Ic SN.
	Compared with the normal type Ic SN 1994I,  SN 1998bw light curves showed to be characterized by a slower initial evolution and a broader light curve
	(peaking at 17 days from the burst, see Fig. 1 in Galama et al. 1998), while the spectra appeared almost featureless, smoothed by the strong blending of many metal lines 
	with very large velocity  ($\sim$ 30,000 km s$^{-1}$ at early phases, Patat et al. 2001). 
         Modeling light curves and spectra, Iwamoto et al. (1998) found that SN 1998bw was best reproduced by an energetic explosion of a C+O star. The initial mass of the progenitor 
         was found to be $\sim 40\,M_\odot $ with a total $^{56}$Ni ejected mass $\sim$0.38--0.48\,$M_\odot$, responsible for the high peak luminosity  (see Table \ref{tab}). 
         The core collapse of such a massive star probably led to the formation of a rapidly rotating black hole (Iwamoto et al. 1998).
         Subsequently, the study of the nebular line profiles (the Fe-dominated blend near 5200\,\AA\ and [O I] $\lambda\lambda$6300 and 6363  lines)
          revealed a  strongly asymmetric explosion geometry, observed relatively off-axis from the jet direction  (under a viewing angle of 15--30 degree; Maeda et al. 2002, 2006).
          As a consequence, taking into account that the highest velocity material is only ejected in a narrow cone  (Maeda et al. 2003) the explosion parameters obtained by using spherically symmetric explosion models, have to be scaled by factors of a few.  \\        
 We had to wait until 2003 for two more clear examples of SN-GRB connections.
The first case was GRB 030329/ SN 2003dh: in contrast to the previous case, the associated GRB had an isotropic emission similar to that of typical cosmological GRBs 
($E_{iso}=1.7\times10^{52}$ erg, see Table \ref{tab}), while SN 2003dh  was highly energetic ($E_k= 4\times 10^{52}$ erg) with a broad light curve (Hjorth et al. 2003).  
The second case was GBR 031203/ SN 2003lw, which resembled the GRB 980425/ SN 1998bw event, with an extremely low energy GRB ($E_{iso}\sim10^{50}$ erg, 
see Table \ref{tab}) and a very luminous and slowly evolving SN ($M_V\sim-19.7$ mag, Malesani et al. 2004). 
The spectra of SN 2003lw were remarkably similar to those of SN 1998bw obtained at comparable epochs. 
From the spectra and light curves modeling, Mazzali et al. (2003) found that the SN 2003dh progenitor  was  a  massive envelope-stripped star with a main sequence mass 
of $\sim$35--40$\,M_\odot$, ejecting  $0.35\,M_\odot$  of $^{56}$Ni. 
Also for SN 2003dh,  the nebular-phase emission lines profile suggested that an aspherical explosion occurred,
maybe viewed almost pole-on, i.e. along its axis direction (Kosugi et al. 2004). 
A more massive progenitor star was deduced for SN 2003lw (40--50 $M_\odot$), leaving behind a black hole remnant (Mazzali et a. 2006b).
No nebular spectra were taken for SN 2003lw, but taking into account its observed properties as well as those of the connected GRB, 
our line of sight was likely lying with an angle from the jet direction in between those of SNe 1998bw and 2003lw.\\
The last case in the sample is  GRB 060218/ SN 2006aj.
GRB 060218 was an unusual low-luminosity and  long-dura\-tion ($T_{90} = 2100$ s) event. 
Its X-ray prompt emission was formed by the typical non-thermal synchrotron component,  coming from the collision of fast-moving 
shells within the GRB jet (e.g. Rees \& Meszaros 1994),  and a thermal one, revealing for the first time the presence of the shock breakout of 
the emerging supernova (Campana et al. 2006; Waxman, Meszaros \& Campana 2007). 
GRB 060218 atypical properties  have been explained as a different nature of the compact central object formed with the explosion: a 
neutron star rather than a black hole (Toma et al. 2007 ; Mazzali et al. 2006a; Fan et al. 2010). 
SN 2006aj displayed a faster spectral evolution than the previous GRB-SN  cases, 
in  agreement with the narrow light curve (Pian et al. 2006; Mazzali et al. 2006a). 
A rapidly evolving light curve is possibly related to a small ejecta mass ($M_{ej}$) and a low total explosion energy ($E_k$).
Indeed from the spectra and light curve modeling, Mazzali et al. (2006a) found that the supernova had a
 $E_k=2\times10^{51}$ erg and only $M_{ej}=2\,M_\odot$, suggesting that it was produced 
by a star whose initial mass was $\sim 20\,M_\odot$. For star of this mass, the central compact object formed in the core collapse is expected to 
be a neutron star rather than a black hole (Mazzali et al. 2006a). \\
The SN-GRB connection thus seems to involve  different physical mechanisms: a "collapsar'' for the more massive stars collapsing to a black hole,
 and a magnetar for the less massive stars (Woosley \& Bloom 2006).

\section{The lastest case of GRB-SN connection: GRB 100316D/ SN 2010bh}

GRB 100316D was detected with the Swift Burst Alert Te\-lescope (BAT) on 2010 March 16.531 UT 
(Stamatikos et al. 2010). It turned out to be an atypical GRB, whose prompt emission was characterized by a very soft spectral 
peak and an extended and slowly decaying flux emission. The estimated total isotropic energy was 
$E_{iso} \sim 4 \times10^{49}$ erg  (Starling et al. 2010).
Similarly to GRB 060218, GRB 100316D had an unusual long duration, lasting  $T_{90} > 1300$ s
and a spectral hardness evolution with a stable and softer spectral shape throughout the 
prompt and late-time emission (Starling et al. 2010). 
Such constant and long X-ray emission could indicate the presence of a great mass of material  whi\-ch feeds the central engine and prolongs its activity (Starling et al. 2010). 
Fan et al. (2010) suggested that the luminosity and the duration of the X-ray emission of GRB 100316D 
could be consistent with the radiation from a magnetar rather than a black hole.
In addition, the X-ray spectrum of  GRB 100316D showed a thermal component 
indicating the presence of the shock break-out of the supernova (Starling et al. 2010).  

\begin{figure*}
\centering
\includegraphics[width=160mm,height=155mm]{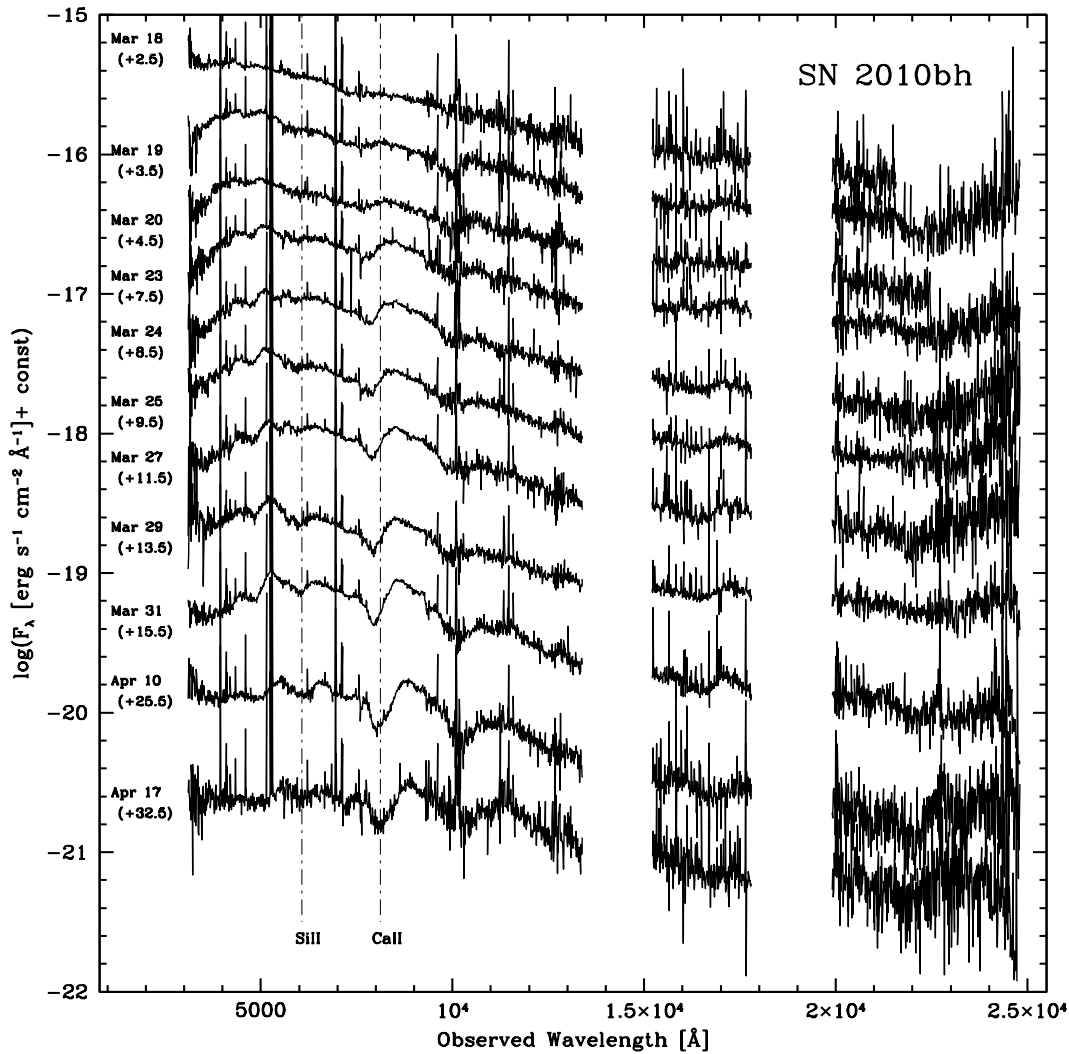}
\caption{Spectral evolution of SN 2010bh over a period of about 2 months obtained with  VLT + X-shooter. Phases are reported in the observed frame from Swift/BAT trigger 
(2010 Mar 16.53, Stamatikos et al. 2010). Spectra have been vertically displaced of an arbitrary quantity for clarity of the plot.}
\label{label_spec}
\end{figure*}

\subsection{SN 2010bh observations with X-shooter}
\begin{figure}
\includegraphics[width=83mm,height=83mm,angle=0]{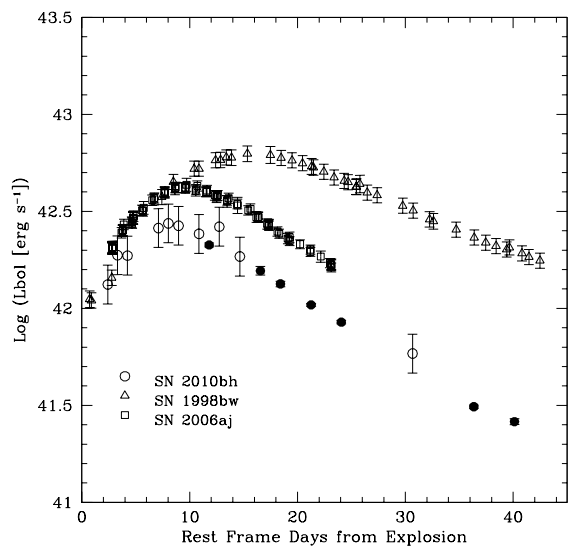}
\caption{SN 2010bh BVRI-bolometric light curve compared to those of SNe 1998bw and 2006aj. SN2010bh pseudo-bolometric luminosities obtained from FORS2 photometry and from XS synthetic photometry are reported with filled and open dots, respectively.}
\label{label_bol}
\end{figure}

The observational campaign with X-shooter started immediately after the GRB 100316D detection. A few days later
the emerging supernova was clearly detected (SN 2010bh; Bufano et al. 2010, Chornock et al. 2010b), confirming the early 
findings by Wiersema et al. (2010)  and Chornock et al. (2010a). 
During the follow up, we collected spectroscopic    data 
with X-shooter for 11 epochs. All observations were performed in ToO mode.\\
 Photometric and spectroscopic data were also collected with VLT/FORS2 and will be presented in a forthcoming comprehensive study 
(Bufano et al. 2011, in prep.). 
The spectroscopic evolution obtained with X-shooter is shown in Fig. \ref{label_spec}.
 Such spectral time sequence  is unprecedented. The first spectrum was ta\-ken only 2.5 days (observer frame) after the burst, 
 while for the previous SN-GRB cases, the earliest SN spectrum was obtained at $\sim$4 days and a week after the 
 burst for SN 2006aj and 1998bw,  respectively. 
 X-shooter allowed us to follow accurately the time evolution of the entire SED  thanks to the extended lambda coverage and spectral resolution.
From Fig.  \ref{label_spec} we can see that  the initial SN 2010bh SED resembles a black body curve with a temperature of $\sim$8500 K.
 Subsequently  the continuum starts to be shaped by broad absorptions, which become deeper and redder with time as an effect of the recession of the photosphere and the drop of the temperature. 
 P-Cygni profiles from the Ca\,II $\lambda$8579 and Si\,II $\lambda$6355 lines are clearly visible around $\sim$8000 \AA\ and $\sim$6000 \AA\ respectively.  \\
Thanks to X-shooter spectra we can also identify a broad absorption due to He\,I $\lambda$10830. Such a detection is important to constrain the nature and the evolutionary state of the progenitor 
prior to the explosion.  At early epochs we can also recognize the optical He\,I line at $\lambda$5876, 
while we do not have a clear detection of the NIR He\,I line at $\lambda$20580, likely due to the 
lower S/N of the spectrum in the NIR range.
 The He\,I $\lambda$10830 line was previously detected only in SN 1998bw spectra, by Patat et al. (2001).\\
 SN 2010bh exploded in a complex galaxy region, and in order to measure SN 2010bh magnitudes we had to apply a template subtraction method to remove the host galaxy flux contamination.
 Since no pre-explosion images of the field were available, we used  late epoch FORS2 images (taken on September 19, 2010) as templates, assuming that the SN had a negligible
  flux contribution.\\
Though generally aimed to spectroscopic analysis, X-shoo\-ter provided important photometric data too. We were able to obtain good photometry with 
X-shooter by using its acquisition camera images.  In this way, we were able to follow the very early  photometric evolution of SN 2010bh, that was not possible to cover with 
VLT/FORS2.  Acquisition R-band images were consistent with the FORS2 photometric system so we could use the same template image to measure the 
SN magnitudes and the same local sequence stars to calibrate the magnitudes from the instrumental to the standard system.
Thus, thanks to X-shooter, we were able to trace the R light curve from the early post-explosion phases (assumed coincident with the GRB trigger; Stamatikos M. et al. 2010) up to
a couple of months later.\\
By using the R magnitudes, we performed the absolute flux calibration of the spectra, allowing us to calculate directly from them the BVRI synthetic magnitudes for each epoch.
In Fig. \ref{label_bol}, the  BVRI-bolometric light curve is shown, where the pseudo-bolometric luminosity obtained from FORS2\\ photometry is reported with filled dots,
while the empty dots show the luminosity computed from X-shooter spectra synthetic magnitudes. We show SNe 1998bw and 2006aj in the same plot for comparison purpose. 
SN 2010bh  has a faster rise to the peak, a fainter peak ($L_{bol}\sim3 \times 10^{42}$ erg\,s$^{-1}$) and a more rapid decline than the other SNe.
This likely means  an high energetic explosion with a small ejected mass and  a small mass of ejected radioactive $^{56}$Ni.
A detailed discussion on the explosion scenario of SN 2010bh will be presented in Bufano et al. (2011, in prep.).

\section{Conclusion}
 We presented SN 2010bh, the lastest SN Type Ib/c observed in connection with a GRB, GRB 100316D. 
 In particular, we showed the important role that VLT plus X-shooter played during the observational
 campaign. Thanks to its extended wavelength coverage, X-shooter allowed us to 
 obtain the SN spectral energy distribution  from the UV to the NIR range in a single shot and follow its evolution with time. 
 In addition, we were able with X-shooter to obtain good photometry by using its acquisition images. Thus we could 
 measure the SN 2010bh magnitudes and, through the absolute flux calibrated spectra, its pseudo-bolometric light curve. 
 Such information is fundamental to probe the explosion scenario and understand the progenitor nature.

\newpage


\end{document}